\documentclass[10pt,conference]{IEEEtran}
\IEEEoverridecommandlockouts
\usepackage[nobreak,space,compress]{cite}
\usepackage{amsmath,amssymb,amsfonts}
\usepackage{algorithmic}
\usepackage{graphicx}
\usepackage{textcomp}
\usepackage{xcolor}
\def\BibTeX{{\rm B\kern-.05em{\sc i\kern-.025em b}\kern-.08em
    T\kern-.1667em\lower.7ex\hbox{E}\kern-.125emX}}

\usepackage{comment}
    
\begin{document}

\title{Task-Aware Reduction for Scalable LLM–Database Systems\\
}

\author{\IEEEauthorblockN{1\textsuperscript{st} Marcus Emmanuel Barnes}
\IEEEauthorblockA{\textit{Faculty of Information} \\
\textit{University of Toronto}\\
Toronto, Canada \\
marcus.barnes@utoronto}
\and
\IEEEauthorblockN{2\textsuperscript{nd} Taher A. Ghaleb }
\IEEEauthorblockA{\textit{Department of Computer Science} \\
\textit{Trent University}\\
Peterborough, Canada \\
taherghaleb@trentu.ca}
\and
\IEEEauthorblockN{3\textsuperscript{rd} Safwat Hassan}
\IEEEauthorblockA{\textit{Faculty of Information} \\
\textit{University of Toronto}\\
Toronto, Canada \\
safwat.hassan@utoronto.ca}
}

\maketitle

\begin{abstract}
Large Language Models (LLMs) are increasingly applied to data-intensive workflows, from database querying to developer observability. Yet the effectiveness of these systems is constrained by the volume, verbosity, and noise of real-world text-rich data such as logs, telemetry, and monitoring streams. Feeding such data directly into LLMs is costly, environmentally unsustainable, and often misaligned with task objectives. Parallel efforts in LLM efficiency have focused on model- or architecture-level optimizations, but the challenge of reducing upstream input verbosity remains underexplored.

In this paper, we argue for treating the token budget of an LLM as an \emph{attention budget} and elevating task-aware text reduction as a first-class design principle for language--data systems. We position input-side reduction not as compression, but as attention allocation: prioritizing information most relevant to downstream tasks.

We outline open research challenges for building benchmarks, designing adaptive reduction pipelines, and integrating token-budget--aware preprocessing into database and retrieval systems. Our vision is to channel scarce attention resources toward meaningful signals in noisy, data-intensive workflows, enabling scalable, accurate, and sustainable LLM--data integration.
\end{abstract}


\vspace{11pt}
\section{Introduction}
\label{sec:Introduction}
\vspace{4pt}

Large Language Models (LLMs) are increasingly embedded into data-intensive workflows, powering natural language interfaces for databases, observability platforms, and developer tools \cite{roy2024exploring,wang2024rcagent}. They enable capabilities such as conversational querying, automated debugging, and intelligent system triage, transforming how users interact with large-scale, heterogeneous data \cite{xu2009detecting,harty2021logging}. As organizations adopt LLMs at the core of data management and software engineering processes, the integration of language and data has become both a research challenge and a practical necessity.

However, the effectiveness of LLMs in these settings is constrained by real-world data. Logs, telemetry streams, and execution traces are verbose, noisy, and dominated by task-irrelevant content \cite{he2021survey,gholamian2021comprehensive}. Continuous Integration (CI) pipelines, for example, may generate thousands of lines of build and test output, much of it boilerplate or repeated status updates \cite{ghaleb2025,moriconi2025ghalogs, ghaleb2023}. Feeding such raw data directly into LLMs inflates inference cost and latency, wastes token budgets, and introduces noise that obscures the underlying signal. These inefficiencies reduce accuracy and usability while amplifying environmental costs \cite{wu2022sustainable,schwartz2020green,Koenig2025,Naumann2011,Penzenstadler2012}.

Prior work has explored log compression, structural parsing, and deduplication, producing gains in storage and indexing efficiency \cite{yao2021improving,zhu2019logzip,he2017drain,du2017spell}. Recent efforts even apply LLMs for anomaly detection and log filtering \cite{qi2023loggpt,huang2024demystifying}. Yet most techniques remain \emph{task-agnostic}: they reduce size without regard for the semantic needs of downstream tasks, often preserving large volumes of irrelevant content that limit effectiveness in token-sensitive workflows.

We argue for a new paradigm: \textbf{task-aware text reduction pipelines} as a first-class component of language--data integration. Rather than indiscriminately compressing or parsing, these pipelines act as intelligent preprocessing layers that prioritize semantic relevance and explicitly treat the token budget of an LLM as an \emph{attention budget}. By foregrounding task relevance, such pipelines promise three benefits: (1) scalability through token efficiency, (2) sustainability by lowering energy and infrastructure costs, and (3) accuracy by focusing attention on the most relevant signals. This paradigm is complementary to model- and architecture-level efficiency work \cite{10.1145/3530811,dao2022flashattention,spector2023staged} and retrieval-augmented generation and indexing approaches \cite{lewis2020retrieval,chen_kg-retriever_2025,salton1983}.

\vspace{8pt}
\noindent\textbf{Contributions.} This paper makes the following contributions:

\begin{itemize}
    \item \textbf{Conceptual reframing:} We introduce input reduction as an \emph{attention allocation problem}, positioning task-aware reduction as a complementary layer alongside model- and architecture-level efficiency methods~\cite{10.1145/3530811,dao2022flashattention,spector2023staged}.

    \vspace{2pt}
    \item \textbf{Research agenda:} We outline open challenges in evaluation benchmarks, adaptive reduction strategies, token-budget--aware indexing, and sustainability metrics~\cite{lewis2020retrieval,chen_kg-retriever_2025,salton1983}, charting a roadmap for scalable and environmentally responsible integration of LLMs with real-world data systems.

    \vspace{2pt}
    \item \textbf{Domain generality:} We highlight opportunities for task-aware reduction beyond software logs, including healthcare~\cite{boll2025distillnote,mehenni_ontology-constrained_2025,oliveira2025development}, the Internet of Things~\cite{an2024iot,sarhaddi2025llms}, and other data-intensive domains where verbosity threatens efficiency and accuracy.
\end{itemize}


\noindent\textbf{Paper Organization.} The rest of this paper is organized as follows.
Section~\ref{sec:Background} provides background and reviews related work on log analysis and the use of LLMs in this context.
Section~\ref{sec:Vision} introduces our vision for task-aware text reduction pipelines as a new abstraction layer for language–data systems.
Section~\ref{sec:ResearchAgenda} outlines our research agenda and discusses key open challenges.
Finally, Section~\ref{sec:Conclusion} concludes the paper and suggests directions for future work.

\section{Background and Related Work}
\label{sec:Background}

Prior work on log analysis has explored compression, structural parsing, and deduplication as strategies to manage the scale of text-rich data. For example, compression techniques improve log storage efficiency \cite{yao2021improving}, while template-based methods such as LogZip \cite{zhu2019logzip}, Drain \cite{he2017drain}, and Spell \cite{du2017spell} reduce structural variability and support downstream processing. Earlier efforts in anomaly detection mined console logs to identify large-scale system problems \cite{xu2009detecting}. These approaches have proven effective for storage and indexing but remain largely task-agnostic, often retaining significant amounts of irrelevant content.

Recent advances have begun to apply LLMs directly to log data. Qi et al. proposed LogGPT for anomaly detection, showing the potential of transformer-based models for log understanding \cite{qi2023loggpt}. More recently, Huang et al. introduced LoFI, a prompt-based method for extracting fault-indicating information from logs \cite{huang2024demystifying}. While promising, these approaches incur high inference costs or focus on specific tasks, rather than providing a general, reusable reduction layer. In contrast, we position \emph{task-aware text reduction} as a general-purpose paradigm for directing scarce attention resources toward semantically relevant tokens.

A parallel line of work in natural language processing has focused on reducing the computational footprint of large models. Surveys on efficient Transformers \cite{10.1145/3530811} and architectural optimizations such as FlashAttention \cite{dao2022flashattention} propose model-level techniques to accelerate inference, while speculative decoding improves token generation efficiency \cite{spector2023staged}. These methods operate primarily at the level of model architecture, complementary to our focus on upstream input reduction as an attention-allocation problem.

Software engineering and sustainability research further highlights the environmental costs of large-scale computation \cite{wu2022sustainable,schwartz2020green,Koenig2025,Naumann2011,Penzenstadler2012,tripp_measuring_2024,rolf2022dataExternalities}. These studies argue for approaches that balance performance with energy efficiency and ecological impact, motivating our exploration of reduction-oriented preprocessing as a sustainability principle.

In database and information retrieval research, retrieval-augmented generation (RAG) grounds LLMs in structured knowledge sources \cite{lewis2020retrieval}, and recent work explores efficient indexing strategies tailored for LLM workloads \cite{chen_kg-retriever_2025}. Classic IR theory on indexing and query optimization \cite{salton1983} provides additional motivation for treating task-aware preprocessing pipelines as first-class components in the language--data stack.

\section{Vision: Task-Aware Text Reduction Pipeline}
\label{sec:Vision}

We envision \textbf{task-aware text reduction pipelines} as a new abstraction layer for language--data systems. These pipelines operate between raw data streams and LLM inference, filtering and restructuring content so that only semantically relevant information is retained. Unlike compression or syntactic parsing \cite{zhu2019logzip,he2017drain,du2017spell}, which aim to improve storage or indexing efficiency, task-aware pipelines are explicitly guided by the needs of downstream tasks such as failure triage, anomaly detection, or query answering.

This paradigm promises three key benefits. First, \emph{scalability}: by reducing token counts before inference, pipelines lower latency and computational overhead. Second, \emph{sustainability}: trimming unnecessary content reduces the carbon footprint of LLM-driven analysis by minimizing redundant computation and data transfer \cite{Koenig2025,Naumann2011,schwartz2020green,rolf2022dataExternalities,tripp_measuring_2024}. Third, \emph{accuracy}: by exposing models only to semantically relevant signals, task-aware reduction can improve the precision and reliability of downstream outputs. In short, we treat the token budget of an LLM as an \emph{attention budget}, and argue that reduction pipelines are essential to aligning scarce attention resources with the signals that matter most.

We propose three design principles for building such pipelines. \textbf{Task relevance first}: retain the information that contributes directly to the diagnostic or analytic objective, while aggressively filtering boilerplate or low-information content. \textbf{Token-budget awareness}: treat inference cost as a resource allocation problem, preserving the tokens that matter most under explicit budget constraints. \textbf{Hybrid structural--semantic reduction}: combine structural cues (e.g., templates, schema metadata, system events) with semantic methods (e.g., embeddings, task-specific prompts) to identify the segments most relevant to the task.

The scope of task-aware reduction extends well beyond Continuous Integration logs. Similar challenges arise in \emph{cloud observability}, where vast telemetry streams contain only sparse anomaly signals; in \emph{system monitoring}, where traces and events are verbose yet repetitive; and in \emph{knowledge graphs}, where rich metadata and contextual annotations often overwhelm reasoning pipelines. Beyond software systems, \emph{healthcare data} provides a striking example: clinical notes and electronic health records are lengthy and often dominated by boilerplate, yet only a subset of the content is relevant for clinical decision-making. Recent work on clinical summarization demonstrates how reduction can preserve diagnostic fidelity while eliminating unnecessary text \cite{boll2025distillnote,mehenni_ontology-constrained_2025,oliveira2025development}. Likewise, in the \emph{Internet of Things}, continuous sensor and telemetry streams generate massive volumes of largely repetitive data, where the challenge lies in surfacing sparse and semantically important anomalies. Surveys and early frameworks highlight the opportunity for reduction and filtering as prerequisites for effective LLM-based reasoning in IoT domains \cite{an2024iot,sarhaddi2025llms}.

Across these settings, task-aware pipelines balance efficiency with fidelity, creating a unifying paradigm across domains from logs to medical notes to sensor data.

In short, task-aware text reduction offers a foundation for rethinking the role of preprocessing in language--data systems. By foregrounding relevance, these pipelines complement model-level efficiency advances and retrieval-based methods, establishing a new paradigm for scalable, sustainable, and accurate LLM--data integration.

\section{Research Agenda and Open Challenges}
\label{sec:ResearchAgenda}

Task-aware text reduction pipelines open a rich set of research opportunities at the intersection of natural language processing, databases, and software engineering. Realizing the full potential of this paradigm requires addressing several open challenges.

\subsection{Automated Relevance Labeling}
Manual annotation of task-relevant content is not scalable across large datasets or domains. Similar scalability challenges were observed in empirical analyses of Continuous Integration (CI) build data ~\cite{ghaleb2023}, which examined the interplay between build durations and breakages across thousands of integration runs and highlighted the difficulty of managing large, noisy datasets. In the context of root cause analysis, annotation is usually performed using manual log analyses to identify patterns associated with different types of errors~\cite{brandt2020logchunks,ghaleb2019noise}. To address scalability challenges, future work should explore automated approaches to relevance labeling, including (1) heuristic rules that capture common signals such as error codes or exceptions, (2) weak supervision that combines noisy labels from multiple sources, and (3) LLM-assisted annotation to bootstrap relevance classifiers with minimal human effort~\cite{qi2023loggpt,huang2024demystifying}. Recent work on root-cause analysis with LLM-based agents also suggests that domain-specific relevance signals can be learned and reused across settings~\cite{roy2024exploring,wang2024rcagent}. Such approaches would enable pipelines that automatically direct scarce attention budgets to the most meaningful tokens.

\subsection{Adaptive Reduction Strategies}
The optimal degree of reduction varies by context: compilation errors, for instance, may tolerate aggressive pruning, whereas sparse telemetry requires more conservative filtering. Designing adaptive pipelines that tailor reduction dynamically by failure type, domain, or query intent is an important direction for ensuring both efficiency and diagnostic fidelity. This may require hybrid approaches that combine learned models with domain-specific heuristics \cite{roy2024exploring,wang2024rcagent} and should be evaluated alongside model-level efficiency methods such as efficient Transformers \cite{10.1145/3530811}, FlashAttention \cite{dao2022flashattention}, and speculative decoding \cite{spector2023staged}. Together, these methods would enable both computation-aware and attention-aware reduction.

\subsection{Integration with Database and IR Systems}
Task-aware reduction should not operate in isolation but integrate with existing data management infrastructures. One opportunity is \emph{token-budget--aware indexing}, where reduced representations become first-class citizens in query engines. Another is hybrid retrieval pipelines that combine traditional IR methods with LLMs, leveraging reduction to shrink the search space and improve response latency. Embedding reduction into query planning itself may unlock new forms of task-aware optimization \cite{salton1983,lewis2020retrieval,chen_kg-retriever_2025}. In each case, the goal is to align attention budgets with the segments most relevant to the query.

\subsection{Sustainability Metrics and Benchmarks}
A key motivation for reduction is sustainability, yet little work quantifies its real-world benefits. Future efforts should develop benchmarks that measure energy consumption, carbon footprint, and cost savings across LLM pipelines with and without reduction. These metrics would help compare techniques, guide system design, and motivate the adoption of reduction as an environmental as well as technical best practice \cite{wu2022sustainable,schwartz2020green,Koenig2025,rolf2022dataExternalities,tripp_measuring_2024}. Such benchmarks would clarify the sustainability impact of directing attention away from redundant tokens.

\subsection{Beyond Logs}
Although software logs highlight the challenge of verbosity, task-aware reduction applies equally to other domains. 

In \emph{telemetry and observability}, pipelines could surface sparse anomaly signals from large monitoring streams. In \emph{system monitoring}, task-aware filters could trim redundant events while preserving causal signals. In \emph{knowledge graphs}, reduction may help preprocess verbose metadata and contextual annotations for efficient reasoning. 

Beyond software engineering, \emph{healthcare data} offers a striking case. Clinical notes and electronic health records (EHRs) are lengthy, redundant, and often dominated by boilerplate, yet only a small subset of the content is critical for clinical decision-making. Recent advances in clinical summarization demonstrate that reduction can preserve essential signals while improving diagnostic support and mitigating hallucination risks \cite{boll2025distillnote,mehenni_ontology-constrained_2025,oliveira2025development}.

Similarly, in the \emph{Internet of Things}, continuous sensor and telemetry streams generate massive volumes of repetitive data, with only rare anomalies or outliers being relevant. Applying task-aware reduction here would enable LLMs to reason effectively over IoT data without being overwhelmed by redundancy. Surveys and prototypes highlight the need for scalable preprocessing and filtering in IoT-LLM integration, pointing to reduction as a prerequisite for real-world deployment \cite{an2024iot,sarhaddi2025llms}. 

Each of these domains poses unique challenges, from domain-specific semantics in clinical text to high-frequency noise in IoT data. Yet all share the same fundamental need: directing limited attention budgets toward the tokens that matter most.

\subsection{Call to the Community}
We call on the community to treat task-aware reduction as a \emph{foundational design principle} for language--data systems. Just as indexing and query optimization transformed relational databases, relevance-driven reduction has the potential to reshape how unstructured and semi-structured data is processed in LLM pipelines. Achieving this vision requires collaboration across natural language processing, databases, and software engineering communities, building on both model-level efficiency advances \cite{10.1145/3530811,dao2022flashattention} and data-level retrieval frameworks \cite{lewis2020retrieval,chen_kg-retriever_2025}.

\section{Conclusion}
\label{sec:Conclusion}

This paper has argued for \textbf{task-aware text reduction pipelines} as a cornerstone of future language--data systems. By foregrounding semantic relevance, these pipelines address a critical gap in current approaches, enabling scalable, accurate, and sustainable integration of LLMs with noisy, text-rich data sources \cite{wu2022sustainable,schwartz2020green,Koenig2025,Naumann2011,Penzenstadler2012,rolf2022dataExternalities,tripp_measuring_2024}. 

We have positioned input reduction not as a storage or compression problem, but as an \emph{attention allocation problem}: treating the token budget of an LLM as an attention budget that must be directed toward the most relevant signals. This shift in perspective opens a research agenda around automated relevance labeling \cite{qi2023loggpt,huang2024demystifying}, adaptive reduction strategies \cite{roy2024exploring,wang2024rcagent}, hybrid structural--semantic techniques, and integration with database and retrieval systems \cite{salton1983,lewis2020retrieval,chen_kg-retriever_2025}. 

The challenges we highlight are not confined to software logs. They extend to healthcare, where clinical notes and electronic health records demand task-aware summarization to support decision-making \cite{boll2025distillnote,mehenni_ontology-constrained_2025,oliveira2025development}, and to the Internet of Things, where continuous sensor streams require filtering to surface rare anomalies \cite{an2024iot,sarhaddi2025llms}. These diverse domains underscore the generality of reduction as a unifying paradigm for noisy, high-volume, text-rich data.

These directions complement model- and architecture-level efficiency methods \cite{10.1145/3530811,dao2022flashattention,spector2023staged}, together defining a multi-layered approach to efficiency. We encourage the community to treat task-aware reduction not as an afterthought, but as a \emph{foundational design principle}—just as indexing transformed relational databases, reduction has the potential to reshape how unstructured and semi-structured data is processed in LLM pipelines. Embracing this principle is a step toward building hybrid, scalable, and sustainable systems that define the next generation of language--data fusion.

\section*{Acknowledgment}
We acknowledge the support of the Natural Sciences and Engineering Research Council of Canada (NSERC), \textbf{[RGPIN-2021-03969]}.

\bibliographystyle{IEEEtran}
\bibliography{IEEEabrv,references}

\begin{thebibliography}{10}
\providecommand{\url}[1]{#1}
\csname url@samestyle\endcsname
\providecommand{\newblock}{\relax}
\providecommand{\bibinfo}[2]{#2}
\providecommand{\BIBentrySTDinterwordspacing}{\spaceskip=0pt\relax}
\providecommand{\BIBentryALTinterwordstretchfactor}{4}
\providecommand{\BIBentryALTinterwordspacing}{\spaceskip=\fontdimen2\font plus
\BIBentryALTinterwordstretchfactor\fontdimen3\font minus \fontdimen4\font\relax}
\providecommand{\BIBforeignlanguage}[2]{{%
\expandafter\ifx\csname l@#1\endcsname\relax
\typeout{** WARNING: IEEEtran.bst: No hyphenation pattern has been}%
\typeout{** loaded for the language `#1'. Using the pattern for}%
\typeout{** the default language instead.}%
\else
\language=\csname l@#1\endcsname
\fi
#2}}
\providecommand{\BIBdecl}{\relax}
\BIBdecl

\bibitem{roy2024exploring}
D.~Roy, X.~Zhang, R.~Bhave, C.~Bansal, P.~Las-Casas, R.~Fonseca, and S.~Rajmohan, ``Exploring {LLM}-based agents for root cause analysis,'' in \emph{Companion Proceedings of the 32nd ACM International Conference on the Foundations of Software Engineering}, 2024, pp. 208--219.

\bibitem{wang2024rcagent}
Z.~Wang, Z.~Liu, Y.~Zhang, A.~Zhong, J.~Wang, F.~Yin, L.~Fan, L.~Wu, and Q.~Wen, ``{RCAgent}: Cloud root cause analysis by autonomous agents with tool-augmented large language models,'' in \emph{Proceedings of the 33rd ACM International Conference on Information and Knowledge Management}, 2024, pp. 4966--4974.

\bibitem{xu2009detecting}
W.~Xu, L.~Huang, A.~Fox, D.~Patterson, and M.~I. Jordan, ``Detecting large-scale system problems by mining console logs,'' in \emph{Proceedings of the ACM SIGOPS 22nd Symposium on Operating Systems Principles}, 2009, pp. 117--132.

\bibitem{harty2021logging}
J.~Harty, H.~Zhang, L.~Wei, L.~Pascarella, M.~Aniche, and W.~Shang, ``Logging practices with mobile analytics: An empirical study on firebase,'' in \emph{2021 IEEE/ACM 8th International Conference on Mobile Software Engineering and Systems (MobileSoft)}.\hskip 1em plus 0.5em minus 0.4em\relax IEEE, 2021, pp. 56--60.

\bibitem{he2021survey}
S.~He, P.~He, Z.~Chen, T.~Yang, Y.~Su, and M.~R. Lyu, ``A survey on automated log analysis for reliability engineering,'' \emph{ACM computing surveys (CSUR)}, vol.~54, no.~6, pp. 1--37, 2021.

\bibitem{gholamian2021comprehensive}
S.~Gholamian and P.~A. Ward, ``A comprehensive survey of logging in software: From logging statements automation to log mining and analysis,'' \emph{arXiv preprint arXiv:2110.12489}, 2021.

\bibitem{ghaleb2025}
T.~Ghaleb, O.~Abduljalil, and S.~Hassan, ``{CI/CD} configuration practices in open-source {Android} apps: An empirical study,'' \emph{ACM Transactions on Software Engineering and Methodology}, 2024.

\bibitem{moriconi2025ghalogs}
F.~Moriconi, T.~Durieux, J.-R. Falleri, R.~Troncy, and A.~Francillon, ``{GHALogs}: Large-scale dataset of {GitHub Actions} runs,'' in \emph{2025 IEEE/ACM 22nd International Conference on Mining Software Repositories (MSR)}.\hskip 1em plus 0.5em minus 0.4em\relax IEEE, 2025, pp. 669--673.

\bibitem{ghaleb2023}
T.~A. Ghaleb, S.~Hassan, and Y.~Zou, ``Studying the interplay between the durations and breakages of continuous integration builds,'' \emph{IEEE Transactions on Software Engineering}, vol.~49, no.~4, pp. 2476--2497, 2023.

\bibitem{wu2022sustainable}
C.-J. Wu, R.~Raghavendra, U.~Gupta, B.~Acun, N.~Ardalani, K.~Maeng, G.~Chang, F.~Aga, J.~Huang, C.~Bai \emph{et~al.}, ``Sustainable {AI}: Environmental implications, challenges and opportunities,'' \emph{Proceedings of machine learning and systems}, vol.~4, pp. 795--813, 2022.

\bibitem{schwartz2020green}
R.~Schwartz, J.~Dodge, N.~A. Smith, and O.~Etzioni, ``Green {AI},'' \emph{Communications of the ACM}, vol.~63, no.~12, pp. 54--63, 2020.

\bibitem{Koenig2025}
C.~K{\"o}nig, D.~J. Lang, and I.~Schaefer, ``Sustainable software engineering: Concepts, challenges, and vision,'' \emph{ACM Transactions on Software Engineering and Methodology}, vol.~34, no.~5, pp. 1--28, 2025.

\bibitem{Naumann2011}
S.~Naumann, D.~Schmidt, M.~Dick, J.~Kern, and J.~M. M{\"u}ller, ``The {GREENSOFT} model: A reference model for green and sustainable software and its engineering,'' \emph{Sustainable Computing: Informatics and Systems}, vol.~1, no.~4, pp. 294--304, 2011.

\bibitem{Penzenstadler2012}
B.~Penzenstadler, V.~Bauer, C.~C. Calero, and X.~Franch, ``Sustainability in software engineering: A systematic literature review,'' in \emph{Proceedings of the 16th International Conference on Evaluation and Assessment in Software Engineering (EASE)}, 2012.

\bibitem{yao2021improving}
K.~Yao, M.~Sayagh, W.~Shang, and A.~E. Hassan, ``Improving state-of-the-art compression techniques for log management tools,'' \emph{IEEE Transactions on Software Engineering}, vol.~48, no.~8, pp. 2748--2760, 2021.

\bibitem{zhu2019logzip}
J.~Liu, J.~Zhu, S.~He, P.~He, Z.~Zheng, and M.~R. Lyu, ``Logzip: Extracting hidden structures via iterative clustering for log compression,'' in \emph{2019 34th IEEE/ACM International Conference on Automated Software Engineering (ASE)}, 2019, pp. 863--873.

\bibitem{he2017drain}
P.~He, J.~Zhu, Z.~He, J.~Li, and M.~R. Lyu, ``Drain: An online log parsing approach with fixed depth tree,'' in \emph{2017 IEEE International Conference on Web Services (ICWS)}.\hskip 1em plus 0.5em minus 0.4em\relax IEEE, 2017, pp. 33--40.

\bibitem{du2017spell}
M.~Du and F.~Li, ``Spell: Streaming parsing of system event logs,'' in \emph{2016 IEEE 16th International Conference on Data Mining (ICDM)}, 2016, pp. 859--864.

\bibitem{qi2023loggpt}
J.~Qi, S.~Huang, Z.~Luan, S.~Yang, C.~J. Fung, H.~Yang, D.~Qian, J.~Shang, Z.~Xiao, and Z.~Wu, ``{LogGPT}: Exploring {ChatGPT} for log-based anomaly detection,'' in \emph{2023 IEEE International Conference on High Performance Computing \& Communications, Data Science \& Systems, Smart City \& Dependability in Sensor, Cloud \& Big Data Systems \& Application (HPCC/DSS/SmartCity/DependSys)}.\hskip 1em plus 0.5em minus 0.4em\relax IEEE, 2023, pp. 273--280.

\bibitem{huang2024demystifying}
J.~Huang, Z.~Jiang, J.~Liu, Y.~Huo, J.~Gu, Z.~Chen, C.~Feng, H.~Dong, Z.~Yang, and M.~R. Lyu, ``Demystifying and extracting fault-indicating information from logs for failure diagnosis,'' in \emph{2024 IEEE 35th International Symposium on Software Reliability Engineering (ISSRE)}.\hskip 1em plus 0.5em minus 0.4em\relax IEEE, 2024, pp. 511--522.

\bibitem{10.1145/3530811}
\BIBentryALTinterwordspacing
Y.~Tay, M.~Dehghani, D.~Bahri, and D.~Metzler, ``Efficient transformers: A survey,'' \emph{ACM Comput. Surv.}, vol.~55, no.~6, Dec. 2022. [Online]. Available: \url{https://doi.org/10.1145/3530811}
\BIBentrySTDinterwordspacing

\bibitem{dao2022flashattention}
T.~Dao, D.~Fu, S.~Ermon, A.~Rudra, and C.~R{\'e}, ``{FlashAttention}: Fast and memory-efficient exact attention with {IO}-awareness,'' \emph{Advances in neural information processing systems}, vol.~35, pp. 16\,344--16\,359, 2022.

\bibitem{spector2023staged}
B.~Spector and C.~Re, ``Accelerating {LLM} inference with staged speculative decoding,'' \emph{arXiv preprint arXiv:2308.04623}, 2023.

\bibitem{lewis2020retrieval}
P.~Lewis, E.~Perez, A.~Piktus, F.~Petroni, V.~Karpukhin, N.~Goyal, H.~K{\"u}ttler, M.~Lewis, W.-t. Yih, T.~Rockt{\"a}schel \emph{et~al.}, ``Retrieval-augmented generation for knowledge-intensive {NLP} tasks,'' \emph{Advances in neural information processing systems}, vol.~33, pp. 9459--9474, 2020.

\bibitem{chen_kg-retriever_2025}
\BIBentryALTinterwordspacing
W.~Chen, T.~Bai, J.~Su, J.~Luan, W.~Liu, and C.~Shi, ``{KG}-{Retriever}: Efficient knowledge indexing for retrieval-augmented large language models,'' May 2025, arXiv:2412.05547 [cs]. [Online]. Available: \url{http://arxiv.org/abs/2412.05547}
\BIBentrySTDinterwordspacing

\bibitem{salton1983}
G.~Salton and M.~J. McGill, \emph{Introduction to Modern Information Retrieval}, ser. McGraw-Hill Computer Science Series.\hskip 1em plus 0.5em minus 0.4em\relax New York, NY: McGraw-Hill, 1983.

\bibitem{boll2025distillnote}
H.~O. Boll, A.~O. Boll, L.~P. Boll, A.~A. Hanna, and I.~Calixto, ``{DistillNote}: {LLM}-based clinical note summaries improve heart failure diagnosis,'' \emph{arXiv preprint arXiv:2506.16777}, 2025.

\bibitem{mehenni_ontology-constrained_2025}
G.~Mehenni and A.~Zouaq, ``Ontology-constrained generation of domain-specific clinical summaries,'' in \emph{Knowledge {Engineering} and {Knowledge} {Management}}, M.~Alam, M.~Rospocher, M.~van Erp, L.~Hollink, and G.~A. Gesese, Eds.\hskip 1em plus 0.5em minus 0.4em\relax Cham: Springer Nature Switzerland, 2025, pp. 382--398.

\bibitem{oliveira2025development}
J.~D. Oliveira, H.~D. Santos, A.~H.~D. Ulbrich, J.~C. Couto, M.~Arocha, J.~Santos, M.~M. Costa, D.~Faccio, F.~O. Tabalipa, and R.~F. Nogueira, ``Development and evaluation of a clinical note summarization system using large language models,'' \emph{Communications Medicine}, vol.~5, no.~1, p. 376, 2025.

\bibitem{an2024iot}
T.~An, Y.~Zhou, H.~Zou, and J.~Yang, ``{IoT-LLM}: Enhancing real-world {IoT} task reasoning with large language models,'' \emph{arXiv preprint arXiv:2410.02429}, 2024.

\bibitem{sarhaddi2025llms}
F.~Sarhaddi, N.~T. Nguyen, A.~Zuniga, P.~Hui, S.~Tarkoma, H.~Flores, and P.~Nurmi, ``{LLMs and IoT}: A comprehensive survey on large language models and the internet of things,'' \emph{Authorea Preprints}, 2025.

\bibitem{tripp_measuring_2024}
``Measuring the energy consumption and efficiency of deep neural networks: An empirical analysis and design recommendations.''

\bibitem{rolf2022dataExternalities}
\BIBentryALTinterwordspacing
E.~Rolf, B.~Packer, A.~Beutel, and F.~Diaz, ``Striving for data-model efficiency: Identifying data externalities on group performance,'' in \emph{Workshop on Trustworthy and Socially Responsible Machine Learning, NeurIPS 2022}, 2022. [Online]. Available: \url{https://openreview.net/forum?id=\_h\_ikjOEGL\_}
\BIBentrySTDinterwordspacing

\bibitem{brandt2020logchunks}
C.~E. Brandt, A.~Panichella, A.~Zaidman, and M.~Beller, ``Logchunks: A data set for build log analysis,'' in \emph{Proceedings of the 17th International Conference on Mining Software Repositories}, 2020, pp. 583--587.

\bibitem{ghaleb2019noise}
T.~A. Ghaleb, D.~A. Da~Costa, Y.~Zou, and A.~E. Hassan, ``Studying the impact of noises in build breakage data,'' \emph{IEEE Transactions on Software Engineering}, vol.~47, no.~9, pp. 1998--2011, 2019.

\end{thebibliography}

\end{document}